\documentclass{pasj01}

\RequirePackage{mathpazo}
\RequirePackage[T1]{fontenc}

\newcounter{author}
\def\authorcount#1#2{\refstepcounter{author}\label{#1}
                     \altaffiltext{\ref{#1}}{#2}}

\begin{document}

\SetRunningHead{A.Imada et al.}{}

\Received{201X/XX/XX}
\Accepted{201X/XX/XX}

\title{OAO/MITSuME Photometry of Dwarf Novae. \\III. CSS130418:174033+414756}

\author{Akira~\textsc{Imada},\altaffilmark{\ref{affil:hamburg}}$^,$\altaffilmark{\ref{affil:kwasan}*}
  Keisuke~\textsc{Isogai},\altaffilmark{\ref{affil:kyoto}}
  Kenshi~\textsc{Yanagisawa},\altaffilmark{\ref{affil:oao}}
  Nobuyuki~\textsc{Kawai}\altaffilmark{\ref{affil:ttec}}
}

\authorcount{affil:hamburg}{Hamburger Sternwarte, Universit\"at Hamburg, Gojenbergsweg 112, D-21029 Hamburg, Germany}

\authorcount{affil:kwasan}{
  Kwasan and Hida Observatories, Kyoto University, Yamashina, Kyoto 607-8471, Japan}
\email{$^*$a\_imada@kusastro.kyoto-u.ac.jp}

\authorcount{affil:kyoto}{
  Department of Astronomy, Kyoto University, Kyoto 606-8502, Japan}

\authorcount{affil:oao}{
  Okayama Astrophysical Observatory, National Astronomical Observatory of Japan, Asakuchi, Okayama 719-0232, Japan}

\authorcount{affil:ttec}{
Department of Physics, Tokyo Institute of Technology, Ookayama 2-12-1, Meguro-ku, Tokyo 152-8551, Japan}


 
\maketitle

\begin{abstract}

We report on multicolour photometry of the short period dwarf nova CSS130418:174033+414756 during the 2013 superoutburst. The system showed an unusually short superhump period with 0.046346(67) d during stage A, which is one of the shortest periods among dwarf novae below the period minimum. We found that the bluest peaks in $g' - I_{\rm c}$ colour variations tend to coincide with the brightness minima of the superhump modulations. We also studied nightly-averaged superhump amplitudes in $g'$, $R_{\rm c}$, and $I_{\rm c}$ bands and found that they have less dependence on wavelength. These properties are likely to be in common with dwarf novae exhibiting superhumps.  We successfully obtained $g' - R_{\rm c}$ and $R_{\rm c} - I_{\rm c}$ colours during the temporal dip. The colour indices were significantly bluer compared with other dips of WZ Sge-type dwarf novae. By using the period of the growing superhumps, we estimated the mass ratio to be $q$ = 0.077(5), which is much larger than the previous study.

\end{abstract}

\section{Introduction}

Cataclysmic variables (CVs) are close binary stars that consist of a primary white dwarf and a secondary star. The secondary star fills its Roche lobe, transferring gas into the primary Roche lobe through the inner Lagrangian point (L1). As a result, an accretion disc is formed around the white dwarf if the magnetic field of the white dwarf is sufficiently weak (for a review, see, e.g., \cite{war95book}; \cite{hel01book}).

CVs are divided into several subclasses according to the overall light curves and underlying physics causing the variations. Dwarf novae are a subclass of CVs. They exhibit outbursts with recurrent time scales of days to years with amplitudes of 2$-$8 mag (for a review, see, e.g., \cite{osa96review}; \cite{las01DIDNXT}). The overall behaviour of dwarf novae is well explained by the thermal limit cycle-instability model of the accretion disc
(\cite{mey81DNoutburst}; \cite{sma84DI}).

Dwarf novae are further divided into several subclasses with respect to
their light curves (\cite{VSNET}; \cite{pdot}). SU UMa-type dwarf novae, one subclass of dwarf novae, show two types of outbursts. One is called normal outburst, whose duration is a few days, and the other is called superoutburst, whose duration is typically 2 weeks or sometimes longer and the amplitude is about 1 mag larger than the normal outburst. During the superoutburst, a rapid rise and slow decline modulations, termed superhumps, are observed. The period of the superhump ($P_{\rm sh}$) is slightly longer than the orbital period of the system ($P_{\rm orb}$). Long-term light curves of SU UMa-type dwarf novae are well reproduced by the combination of the thermal and tidal instability model (\cite{osa89suuma}; \cite{osa13v1504cygKepler}). The light curves of superhumps are understood as tidal dissipation of the eccentrically-deformed precessing accretion disc
(\cite{whi88tidal}; \cite{hir90SHexcess}).

Recent extensive photometry during superoutbursts has revealed that the light curves of SU UMa-type dwarf novae show a wide variety (e.g., \cite{pdot}). WZ Sge-type dwarf novae, one subclass of SU UMa-type dwarf novae, are one of the most important systems among them, in terms of researching for various physics of the accretion discs, as well as inspection of evolutionary theories of CVs driven by the gravitational wave radiation \citep{kat15wzsge}. The main characteristics of WZ Sge-type dwarf novae is that (1) they have a long quiescence, typically exceeding a decade (\cite{nog97alcom};
\cite{kat01hvvir}) (2) the maximum magnitude of the superoutburst is more than 6 mag brighter than the quiescent magnitude (\cite{how95TOAD}; \cite{ish01rzleo}), (3) an early stage of the superoutburst show double-peaked modulations called early superhumps (\cite{osa02wzsgehump}; \cite{kat02wzsgeESH}), (4) some of them show rebrightening(s) after the end of the superoutburst plateau (\cite{pat02wzsge}; \cite{kat04egcnc}), and (5) they are absent of normal outbursts (For a review, see \cite{kat15wzsge}).

It is well known that the period of superhumps changes over the course of the superoutburst \citep{pdot}. Our understanding of the superhump period change has been significantly improved over the past decade, mainly by statistical studies of superhumps conducted by VSNET \citep{VSNET}. VSNET (= Variable Star NETwork) is a world-wide amateur-professional network of researchers in variable stars, particularly in transient objects such as cataclysmic variables \citep{pdot}, black-hole binaries \citep{v404}, supernovae \citep{oky93sn1993j}, and gamma-ray bursts \citep{uem03grb030329nat}.\footnote{www.kusastro.kyoto-u.ac.jp/vsnet} According to \citet{pdot}, the evolution of the superhump period consists of three stages: an early stage with a longer and constant period (stage A), a middle stage with positive $P_{\rm dot}$ = ${\dot P}/P$ (stage B), and a late stage with a shorter and constant period (stage C). Although the underlying physics in each stage is not clear, it has been gradually accepted that stage A superhump period is associated with the dynamical precession rate at the 3:1 resonance radius, based on the observations of eclipsing SU UMa-type dwarf novae (\cite{osa13v344lyrv1504cyg}; \cite{kat13qfromstageA}).

Although extensive light curves of superoutbursts have been acquired, the majority of CCD photometry were performed without filters. Recent multicolour observations during superoutbursts have revealed a wide variety of colour variations. For example, \citet{uem08j1021} performed simultaneous optical and near-infrared observations of a WZ Sge-type dwarf nova IK Leo during the 2006 superoutburst and found a significant $K_{\rm s}$ excess during the rebrightening. \citet{uem08j1021} noted that the near-infrared activities provide evidence for the presence of mass reservoir at the outer region of the accretion disc. \citet{mat09v455and} performed multicolour observations during the 2007 superoutburst of V455 And, in which they found that the bluest peak of colour variations of superhumps are prior to their maximum brightness by a phase of 0.15. Regarding the underlying mechanism of colour variations, \citet{mat09v455and} suggested that the emitting size of the superhump light source changes as a result of viscous heating and cooling of the accretion disc, which may be observed as the difference in the maximum phase between the magnitude and colour. On the other hand, \citet{iso15ezlyn} found that the bluest peak of $g' - i'$ colour correspond to the brightness minimum of superhumps during the 2010 superoutburst of EZ Lyn. Similar results were obtained in the WZ Sge-type dwarf novae HV Vir, OT J012059.6$+$325545, and SSS J122221.7$-$311525 (\cite{nak13j0120}; \cite{neu17j1222}; \cite{ima18hvvirj0120}). \citet{iso15ezlyn} noted that the different results between V455 And and EZ Lyn may be caused by the different disc radius between them, based on their photometric and spectroscopic observations. \citet{neu17j1222} studied a relation between superhumps and colour variations according to each superhump stage and found that stage A superhumps showed weaker colour variations compared with stage B ones. \citet{ima18hvvirj0120} suggest that drastic colour variations during stage B may attribute to the predominance of the pressure effect. Regarding the post-superoutburst stage, \citet{neu17j1222} found that the superhump maxima coincide with the bluest peak of $B - I$ colour variations. In order to quantitatively study the colour variations of superhumps, further multicolour photometry is indispensable.

Recent studies have shown that the histogram of the orbital period distribution of dwarf novae has a cutoff around 78 min, below which dwarf novae hardly exists (\cite{gan09SDSSCVs}; \cite{uem10shortPCV}; \cite{kni11CVdonor}). Nevertheless, several systems indeed have their orbital periods far below 78 min \citep{bre12j1122}. Such systems include e.g., V485 Cen (\cite{aug93v485cen}; \cite{ole97v485cen}), EI Psc (\cite{wei01j2329}; \cite{tho02j2329}; \cite{uem02j2329}), SBS 1108+574 (\cite{car13sbs1108}; \cite{lit13sbs1108}; \cite{ken15sbs1108}), and OV Boo (\cite{szk05SDSSCV4}; \cite{lit07j1507}; \cite{pat08j1507}). Theoretical studies suggest that dwarf novae can evolve below the period minimum if the system contains an evolved secondary \citep{pod03amcvn}. Such systems can provide evidence for helium enrichment in their spectra \citep{tho02j2329}, together with N-enhanced and C-depleted features \citep{gan03eipscv396hyabzumaeycyg}.

On 2013 April 18.46, Catalina Real Transient Survey (CRTS, \cite{CRTS}) detected an eruptive object (CSS130418:174033+414756, hereafter, CSS J1740) with the magnitude of ${\sim}$ 14. The All Sky Automated Survey for SuperNovae (ASAS-SN, \cite{ASASSN}) also detected the outburst on 2013 April 19.53 with $V$ = 12.7 \citep{atel4999}. It turned out that the eruption was a superoutburst of a dwarf nova ([vsnet-alert 15639]). Surprisingly, the system showed an unusually short early superhump period with 0.04503(1) ([vsnet-alert 15639]), which is far below the observational period minimum. After this report, we started simultaneous $g'$, $R_{\rm c}$, and $I_{\rm c}$ photometry using the MITSuME telescope located in the Okayama Astrophysical Observatory (OAO).\footnote{www.oao.nao.ac.jp} This system underwent further outbursts in 2007 May and 2014 August ([vsnet-alert 17610]). Here we report on multicolour photometry and analyses during the 2013 superoutburst of CSS J1740.

\section{Observations}

Time-resolved CCD photometry was performed from 2013 April 26 to 2013 May 23 using MITSuME 50cm-telescope located in Okayama Astrophysical Observatory (OAO). MITSuME 50cm-telescope is a robotic telescope that can acquire $g'$, $R_{\rm c}$, and $I_{\rm c}$ images simultaneously by using two dichroic mirrors and three CCD cameras, which allows us to study colour variations of variable stars (For a detail description, see \cite{3me}). We list the log of our observations in table \ref{log}. All data were obtained with the exposure time of 30 sec. The total datapoints of our observations amount to 6524, which is sufficient for period and colour studies of the superoutburst.

After de-biasing, dark-subtracting and flat-fielding the images
with the standard procedure, the data were processed with aperture
photometry using IRAF/daophot.\footnote{IRAF (Image Reduction and
  Analysis Facility) is distributed by the National Optical Astronomy
  Observatories, which is operated by the Association of Universities
  for Research in Astronomy, Inc., under cooperative agreement with
  National Science Foundation.} We performed differential photometry using the star with RA 17:40:11.00, Dec +41:48:18.22, g' = 12.47(6), r' = 12.25(2), and i' = 12.17(3). The constancy of the comparison star is checked by nearby stars in the same field. Because the exact magnitudes of $R_{\rm c}$ and $I_{\rm c}$ of the comparison star are unknown, we converted the SDSS magnitudes to $R_{\rm c}$ and $I_{\rm c}$ given by \citet{2002AJ....123.2121S} as follows:

\begin{equation}
V = g' - 0.55(g' - r') - 0.03
\end{equation}
\begin{equation}
V - R = 0.59(g' - r') + 0.11
\end{equation}
\begin{equation}
R - I = 1.00(r' - i') + 0.21
\end{equation}

Using the above equations, we adopt $g'$ = 12.467(37), $R_{\rm c}$ = 12.077(30), and $I_{\rm c}$ = 11.793(37) as the magnitudes of the comparison star. Barycentric correction was made before the following analyses.

\section{Results and Discussion}

\subsection{Light curve}

\begin{figure}
\begin{center}
\FigureFile(80mm,80mm){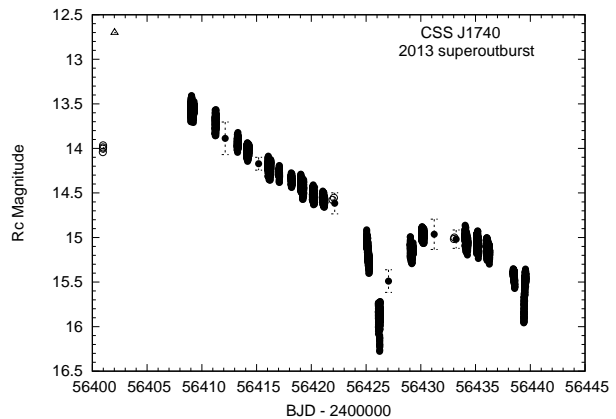}
\end{center}
\caption{The light curve of the 2013 superoutburst of CSS J1740. The abscissa denotes BJD $-$ 2400000, while the ordinate denotes $R_{\rm c}$ magnitude. We also plot $V$ band data of the CRTS (open circles), the ASAS-SN (open triangle), and a part of the AAVSO archival light curve (on BJD 2456438-39). A temporal dip was observed on BJD 2456425, after which the light curve showed a long-lasting rebrightening. A short-term rebrightning was caught on BJD 2455639.}
\label{lc_j1740}
\end{figure}

\begin{figure}
\begin{center}
\FigureFile(80mm,80mm){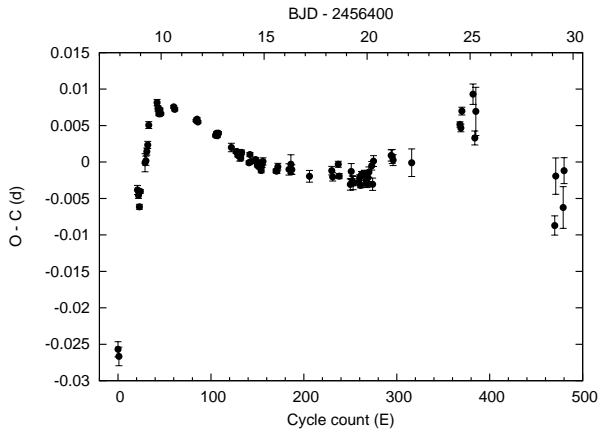}
\end{center}
\caption{$O - C$ diagram of superhump maxima. We used table \ref{o-ctable} and equation (4) to draw the diagram. Stage A can be clearly recognized between BJD 2456408$-$09. This diagram also indicates that the stage B-C transition occurred on BJD 2456425.}
\label{o-c}
\end{figure}

\begin{figure}
\begin{center}
\FigureFile(80mm,80mm){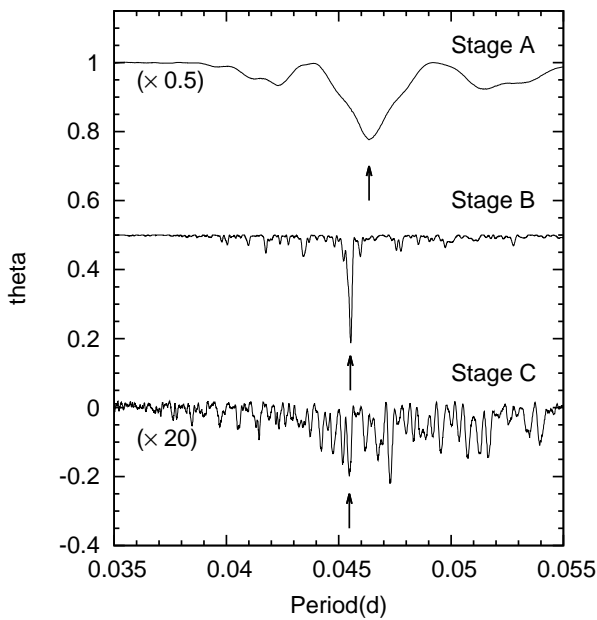}
\end{center}
\caption{Theta diagrams of each stage. The best estimated periods were 0.046346(67) d (stage A), 0.045515(29) d (stage B), and 0.045473(25) d (stage C), respectively, which are marked with the arrows. For better visualization, the theta diagrams of stage A and stage C are changed from the original sizes.}
\label{pdm}
\end{figure}

\begin{table}
\caption{Log of observations of CSS J1740 using MITSuME telescope.}
\begin{center}
\begin{tabular}{cccc}
\hline\hline
Date & BJD(start)$^*$ & BJD(end)$^*$ & N$^{\dagger}$ \\
\hline
2013 Apr. 26 & 56408.9927 & 56409.3060 & 589 \\ 
2013 Apr. 28 & 56411.2084 & 56411.3055 & 138 \\ 
2013 Apr. 29 & 56412.1268 & 56412.1340 & 10 \\ 
2013 Apr. 30 & 56413.2322 & 56413.3033 & 161 \\ 
2013 May. 1 & 56414.0946 & 56414.3035 & 389 \\ 
2013 May. 2 & 56415.1750 & 56415.1842 & 16 \\ 
2013 May. 3 & 56416.0347 & 56416.3020 & 423 \\ 
2013 May. 4 & 56417.0550 & 56417.0942 & 88 \\ 
2013 May. 5 & 56418.1552 & 56418.2397 & 182 \\ 
2013 May. 6 & 56419.0304 & 56419.2997 & 530 \\ 
2013 May. 7 & 56420.1013 & 56420.2983 & 361 \\ 
2013 May. 8 & 56421.0425 & 56421.2169 & 340 \\ 
2013 May. 9 & 56422.0385 & 56422.2542 & 304 \\ 
2013 May. 12 & 56425.0514 & 56425.2945 & 427 \\ 
2013 May. 13 & 56426.1250 & 56426.2916 & 304 \\ 
2013 May. 14 & 56427.0457 & 56427.0554 & 21 \\ 
2013 May. 16 & 56429.0365 & 56429.2921 & 474 \\ 
2013 May. 17 & 56430.0848 & 56430.2928 & 385 \\ 
2013 May. 18 & 56431.1932 & 56431.2931 & 17 \\ 
2013 May. 20 & 56433.1922 & 56433.2626 & 91 \\ 
2013 May. 21 & 56434.0436 & 56434.2907 & 454 \\ 
2013 May. 22 & 56435.0930 & 56435.2924 & 345 \\ 
2013 May. 23 & 56436.0103 & 56436.2882 & 475 \\ 
\hline
\multicolumn{4}{l}{$^*$BJD$-$2400000 $^{\dagger}$ Number of exposure.}
\end{tabular}
\end{center}
\label{log}
\end{table}

\begin{longtable}{cccccccccccc}
\caption{Timings of superhump maxima.}\label{o-ctable}
\hline\hline
E & Max$^*$ & Error & O - C$^{\dagger}$ & N$^{\ddagger}$ & Source${\S}$ & E & Max$^*$ & Error & O - C$^{\ddagger}$ & N$^{\ddagger}$ & Source${\S}$ \\
\hline
\endhead
\endfoot
0 & 56407.6657 & 0.0010 & -0.0257 & 80 & A & 156 & 56414.7972 & 0.0005 & 0.0001 & 54 & A \\
1 & 56407.7103 & 0.0013 & -0.0267 & 79 & A & 170 & 56415.4335 & 0.0002 & -0.0012 & 90 & A \\
21 & 56408.6441 & 0.0006 & -0.0038 & 112 & A & 171 & 56415.4791 & 0.0003 & -0.0012 & 87 & A \\
22 & 56408.6889 & 0.0004 & -0.0046 & 158 & A & 172 & 56415.5252 & 0.0005 & -0.0006 & 69 & A \\
23 & 56408.7329 & 0.0003 & -0.0061 & 154 & A & 184 & 56416.0714 & 0.0008 & -0.0010 & 79 & M \\
24 & 56408.7805 & 0.0003 & -0.0040 & 156 & A & 186 & 56416.1633 & 0.0013 & -0.0003 & 45 & M \\
29 & 56409.0122 & 0.0012 & -0.0001 & 77 & M & 187 & 56416.2081 & 0.0007 & -0.0010 & 76 & M \\
30 & 56409.0580 & 0.0006 & 0.0002 & 82 & M & 206 & 56417.0726 & 0.0008 & -0.0020 & 77 & M \\
31 & 56409.1048 & 0.0005 & 0.0014 & 82 & M & 230 & 56418.1665 & 0.0006 & -0.0012 & 63 & M \\
32 & 56409.1513 & 0.0005 & 0.0023 & 81 & M & 231 & 56418.2112 & 0.0006 & -0.0020 & 76 & M \\
33 & 56409.1996 & 0.0005 & 0.0051 & 80 & M & 237 & 56418.4863 & 0.0004 & -0.0003 & 83 & A \\
42 & 56409.6126 & 0.0004 & 0.0082 & 61 & A & 238 & 56418.5302 & 0.0004 & -0.0019 & 89 & A \\
43 & 56409.6573 & 0.0004 & 0.0073 & 79 & A & 250 & 56419.0756 & 0.0008 & -0.0031 & 76 & M \\
44 & 56409.7022 & 0.0002 & 0.0066 & 78 & A & 251 & 56419.1230 & 0.0011 & -0.0013 & 75 & M \\
45 & 56409.7483 & 0.0002 & 0.0072 & 79 & A & 252 & 56419.1672 & 0.0007 & -0.0026 & 80 & M \\
46 & 56409.7933 & 0.0002 & 0.0066 & 77 & A & 253 & 56419.2125 & 0.0009 & -0.0029 & 78 & M \\
60 & 56410.4319 & 0.0002 & 0.0076 & 87 & A & 259 & 56419.4859 & 0.0005 & -0.0028 & 85 & A \\
61 & 56410.4771 & 0.0002 & 0.0072 & 87 & A & 260 & 56419.5321 & 0.0004 & -0.0021 & 81 & A \\
84 & 56411.5232 & 0.0002 & 0.0057 & 87 & A & 261 & 56419.5766 & 0.0003 & -0.0032 & 88 & A \\
85 & 56411.5690 & 0.0002 & 0.0059 & 82 & A & 262 & 56419.6236 & 0.0004 & -0.0017 & 65 & A \\
86 & 56411.6141 & 0.0002 & 0.0055 & 90 & A & 265 & 56419.7604 & 0.0004 & -0.0016 & 73 & A \\
105 & 56412.4777 & 0.0002 & 0.0036 & 87 & A & 266 & 56419.8060 & 0.0004 & -0.0016 & 76 & A \\
106 & 56412.5236 & 0.0002 & 0.0040 & 89 & A & 267 & 56419.8506 & 0.0004 & -0.0025 & 76 & A \\
107 & 56412.5688 & 0.0002 & 0.0036 & 92 & A & 268 & 56419.8955 & 0.0004 & -0.0031 & 77 & A \\
108 & 56412.6147 & 0.0002 & 0.0040 & 91 & A & 269 & 56419.9420 & 0.0005 & -0.0021 & 74 & A \\
122 & 56413.2504 & 0.0006 & 0.0020 & 82 & M & 270 & 56419.9883 & 0.0007 & -0.0014 & 77 & A \\
127 & 56413.4776 & 0.0002 & 0.0014 & 89 & A & 273 & 56420.1258 & 0.0006 & -0.0005 & 74 & M \\
128 & 56413.5231 & 0.0002 & 0.0014 & 91 & A & 274 & 56420.1689 & 0.0008 & -0.0031 & 80 & M \\
129 & 56413.5681 & 0.0002 & 0.0009 & 86 & A & 275 & 56420.2176 & 0.0007 & 0.0001 & 76 & M \\
130 & 56413.6140 & 0.0003 & 0.0012 & 151 & A & 294 & 56421.0838 & 0.0008 & 0.0009 & 70 & M \\
131 & 56413.6591 & 0.0007 & 0.0008 & 80 & A & 295 & 56421.1292 & 0.0010 & 0.0008 & 77 & M \\
132 & 56413.7045 & 0.0003 & 0.0006 & 79 & A & 296 & 56421.1743 & 0.0007 & 0.0003 & 76 & M \\
133 & 56413.7508 & 0.0003 & 0.0013 & 79 & A & 316 & 56422.0849 & 0.0019 & -0.0001 & 69 & M \\
141 & 56414.1137 & 0.0003 & -0.0001 & 162 & M & 368 & 56424.4587 & 0.0004 & 0.0052 & 168 & A \\
142 & 56414.1604 & 0.0003 & 0.0010 & 156 & M & 369 & 56424.5038 & 0.0005 & 0.0047 & 164 & A \\
143 & 56414.2050 & 0.0003 & 0.0001 & 134 & M & 370 & 56424.5516 & 0.0005 & 0.0070 & 166 & A \\
148 & 56414.4330 & 0.0003 & 0.0003 & 85 & A & 382 & 56425.1005 & 0.0014 & 0.0093 & 75 & M \\
149 & 56414.4783 & 0.0002 & 0.0001 & 88 & A & 384 & 56425.1856 & 0.0010 & 0.0033 & 77 & M \\
150 & 56414.5234 & 0.0003 & -0.0004 & 91 & A & 385 & 56425.2348 & 0.0033 & 0.0069 & 78 & M \\
151 & 56414.5687 & 0.0002 & -0.0006 & 89 & A & 470 & 56429.0909 & 0.0013 & -0.0087 & 78 & M \\
152 & 56414.6144 & 0.0003 & -0.0005 & 161 & A & 471 & 56429.1432 & 0.0025 & -0.0019 & 81 & M \\
153 & 56414.6598 & 0.0004 & -0.0006 & 79 & A & 479 & 56429.5033 & 0.0029 & -0.0062 & 49 & A \\
154 & 56414.7048 & 0.0003 & -0.0012 & 80 & A & 480 & 56429.5539 & 0.0018 & -0.0012 & 49 & A \\
155 & 56414.7512 & 0.0004 & -0.0003 & 77 & A & & & & & & \\
\hline
\multicolumn{12}{l}{$^*$BJD-2400000.}\\
\multicolumn{12}{l}{$^{\dagger}$Against max = 2456407.6914 + 0.045549 $E$} \\
\multicolumn{12}{l}{$^{\ddagger}$Total datapoints used to determine the maximum.}  \\
\multicolumn{12}{l}{${\S}$A:AAVSO M:MITSuME.} \\
\end{longtable}

\begin{table}
\caption{Values of the phase difference between superhumps and colour variations.}
\begin{center}
\begin{tabular}{lccc}
\hline\hline
Object & Stage A & Stage B & References \\
\hline
CSS J1740 & 0.5 & 0.3 & this work \\
V455 And & - & 0.15 & 1 \\
EZ Lyn & - & 0.3-0.5 & 2 \\
SSS J1222 & 0.3 & 0.5 & 3 \\
HV Vir & 0.4 & 0.5 & 4 \\
OT J0120 & 0.5 & 0.2 & 4 \\
\hline
\multicolumn{4}{l}{SSS J1222 = SSS J122221.7$-$311525.} \\
\multicolumn{4}{l}{OT J0120 = OT J012059.6+325545.} \\
\multicolumn{4}{l}{1:\citet{mat09v455and}.} \\
\multicolumn{4}{l}{2:\citet{iso15ezlyn}.} \\
\multicolumn{4}{l}{3:\citet{neu17j1222}.} \\
\multicolumn{4}{l}{4:\citet{ima18hvvirj0120}.} \\
\end{tabular}
\end{center}
\label{offset}
\end{table}

\begin{table}
\caption{$g' - R_{\rm c}$ and $R_{\rm c} - I_{\rm c}$ colours during the dips.}
\begin{center}
\begin{tabular}{cccc}
\hline\hline
Object & $g' - R_{\rm c}$ & $R_{\rm c} - I_{\rm c}$ & References \\
\hline
CSS J1740 & 0.02 & 0.20 & this work \\
EG Cnc & 0.49 & 0.40 & 1 \\
WZ Sge & 0.41 & 0.33 & 2 \\
\hline
\multicolumn{4}{l}{1:\citet{pat98egcnc}. 2:\citet{how04wzsge}}
\end{tabular}
\end{center}
\label{dipcol}
\end{table}

Figure \ref{lc_j1740} shows $R_{\rm c}$ light curve of our observations. Also shown are $V$ band data obtained with the CRTS, ASAS-SN and a part of the AAVSO archive.\footnote{https://www.aavso.org/data-download} In combination with our data and the initial detection by the CRTS and ASAS-SN, the plateau stage lasted for ${\sim}$ 23 d. On BJD 2456425, the light curve entered a rapid fading stage and showed a temporal dip on BJD 2456426 with the magnitude faded by ${\sim}$ 1.5 mag with the duration as short as 1 d. Such a dip is remarkably similar to that observed in a well-known WZ Sge-type dwarf nova AL Com (\cite{nog97alcom}; \cite{kim16alcom}). On BJD 2456427, the magnitude quickly recovered from the dip and CSS J1740 entered a long-lasting rebrightening stage. The profile of the light curve during the rebrightening stage is characteristic of type-A one introduced by \citet{ima06tss0222} and \citet{kat15wzsge}. On BJD 2456438, CSS J1740 faded rapidly with a rate of 0.94(5) mag/d, after which a short rising was observed on BJD 2456439 with a rate of $-$2.58(7) mag/d. This is a rare phenomenon for type-A rebrightenings, but a similar light curve was observed during the 2015 superoutburst of the short-period WZ Sge-type dwarf nova ASASSN-15po \citep{nam17}.

\subsection{Superhump period}

\begin{figure*}
\begin{center}
\FigureFile(40mm,40mm){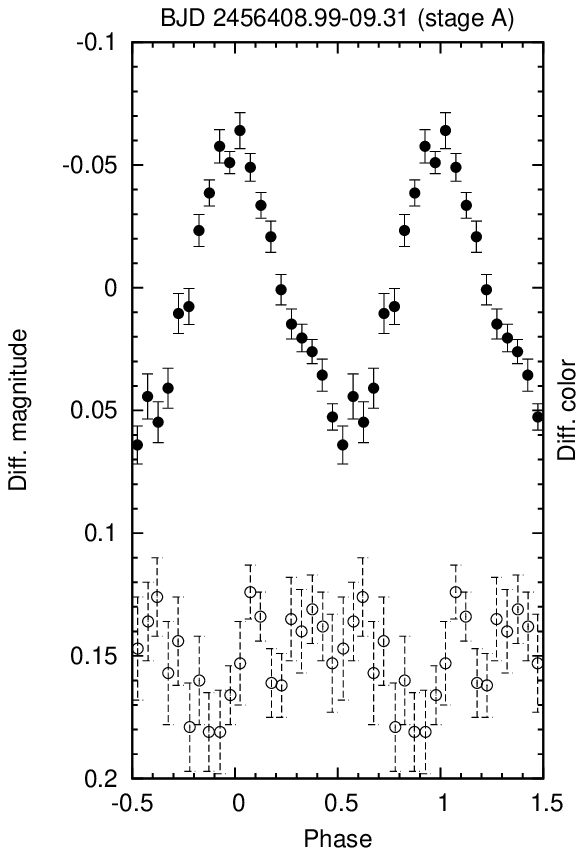}
\FigureFile(40mm,40mm){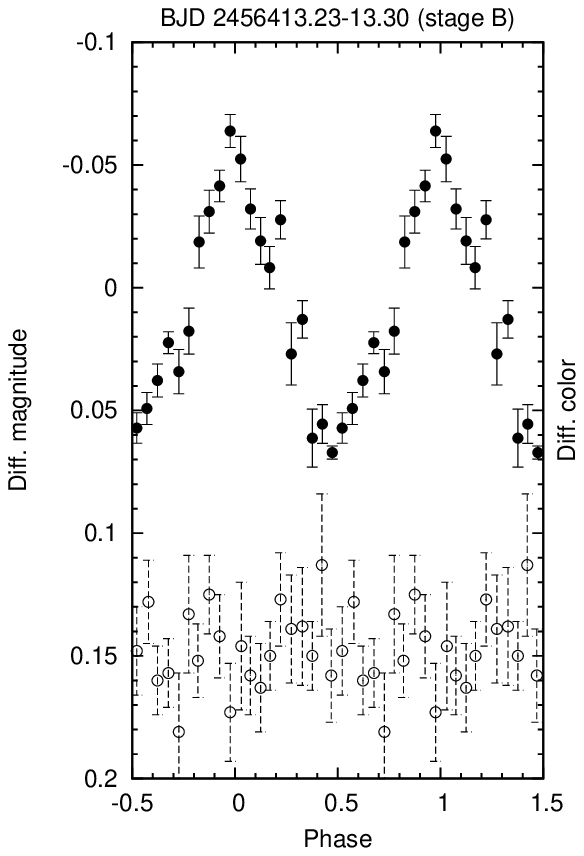}
\FigureFile(40mm,40mm){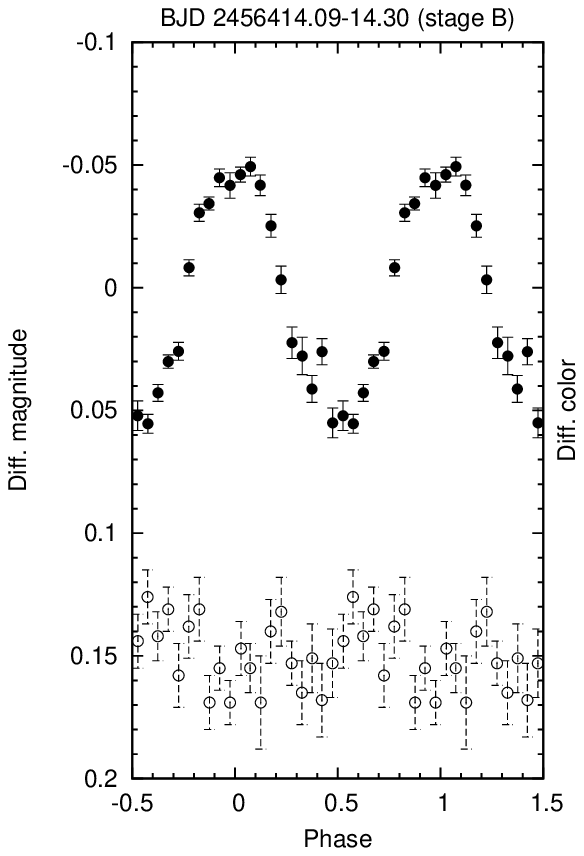}
\FigureFile(40mm,40mm){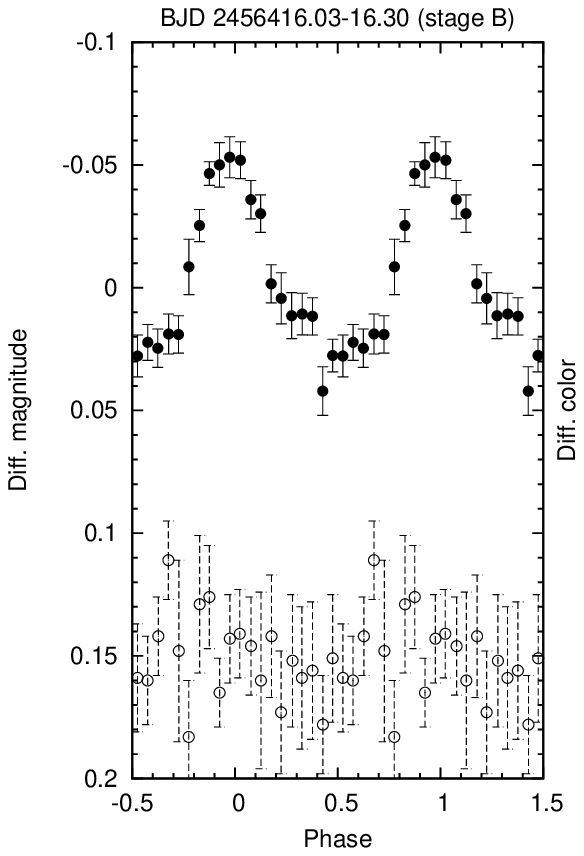}
\FigureFile(40mm,40mm){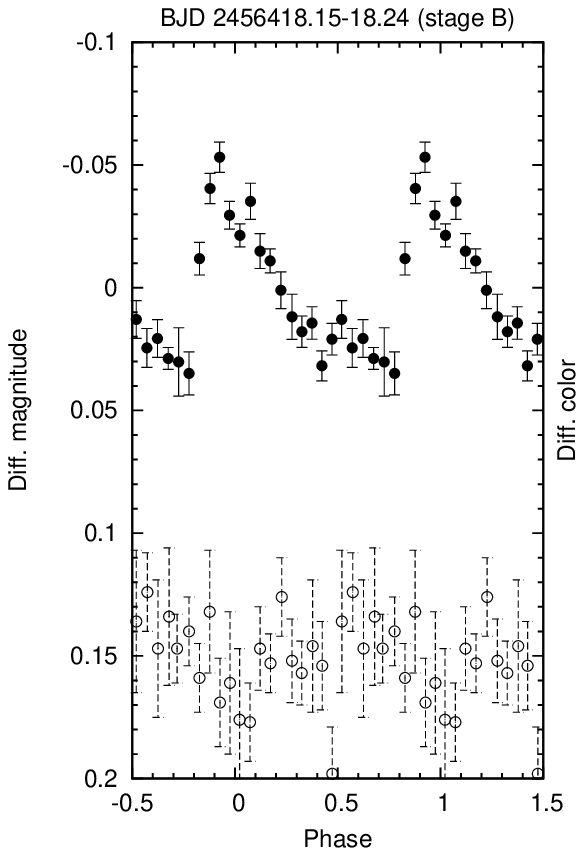}
\FigureFile(40mm,40mm){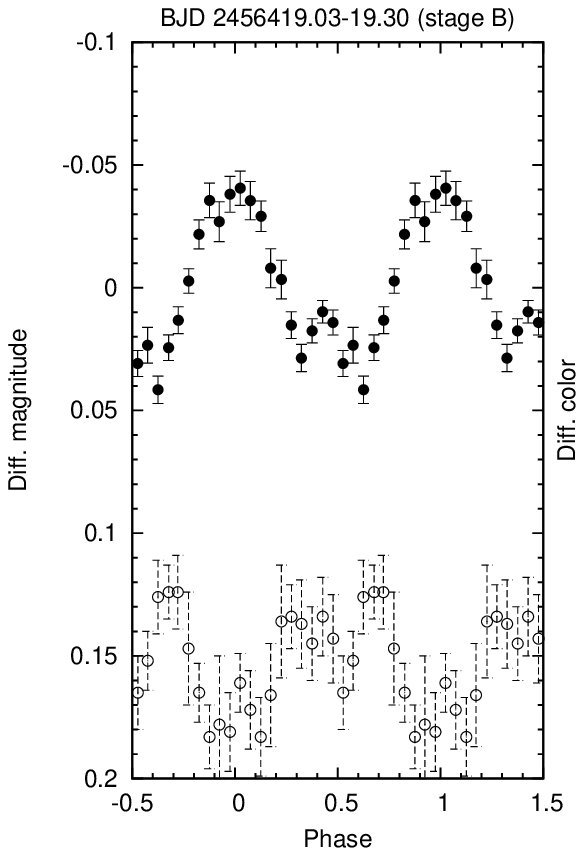}
\FigureFile(40mm,40mm){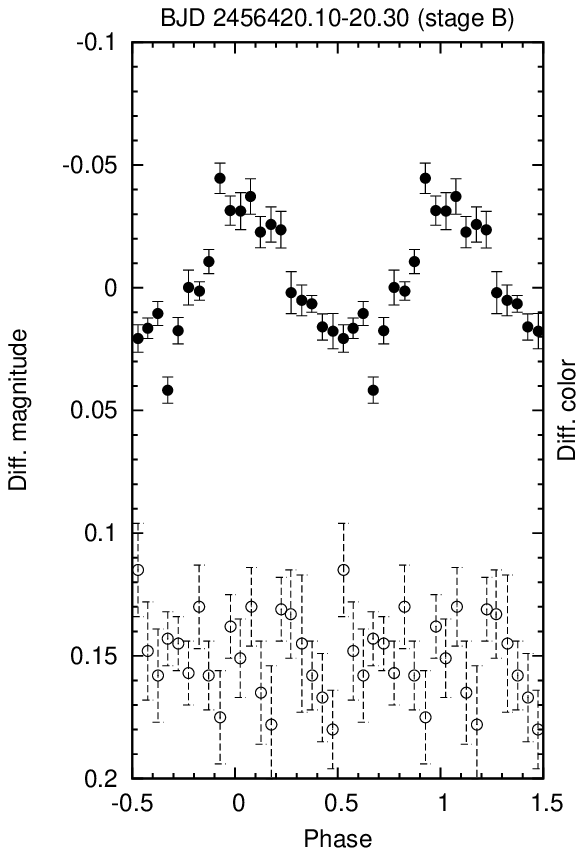}
\FigureFile(40mm,40mm){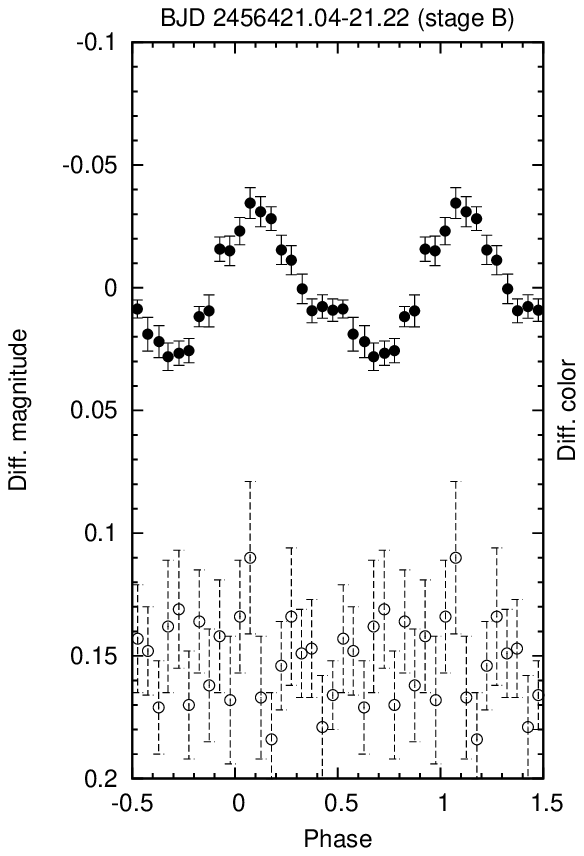}
\FigureFile(40mm,40mm){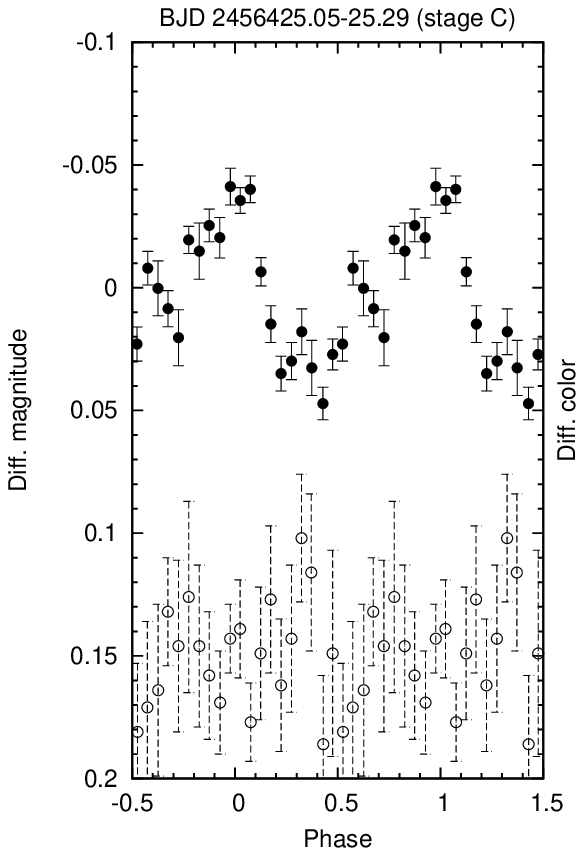}
\FigureFile(40mm,40mm){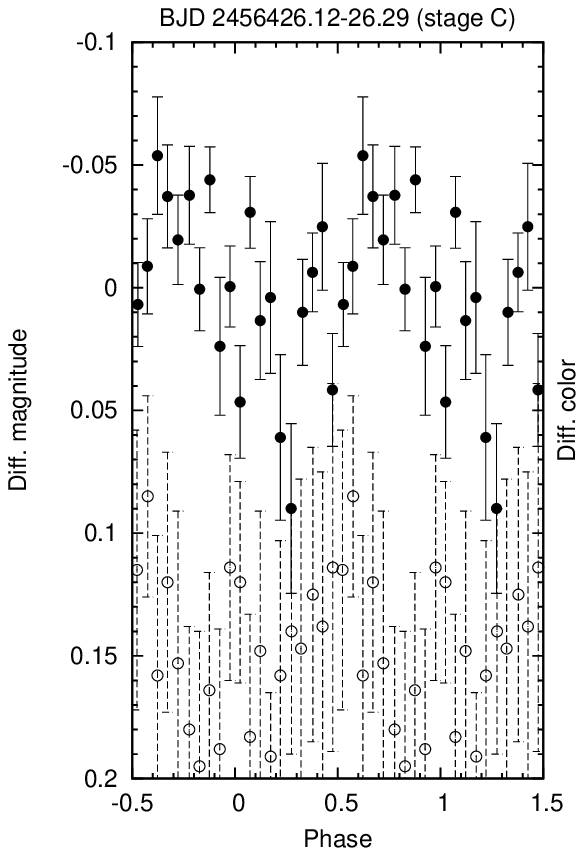}
\FigureFile(40mm,40mm){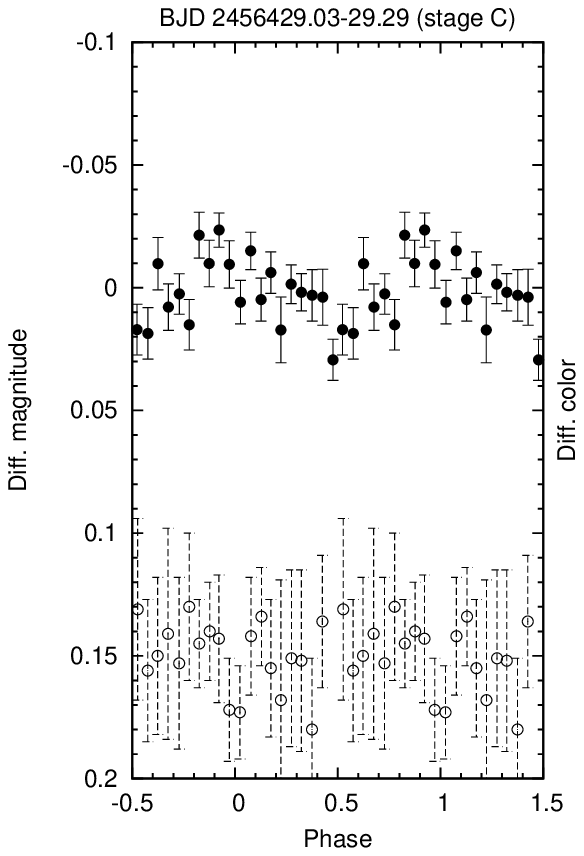}
\FigureFile(40mm,40mm){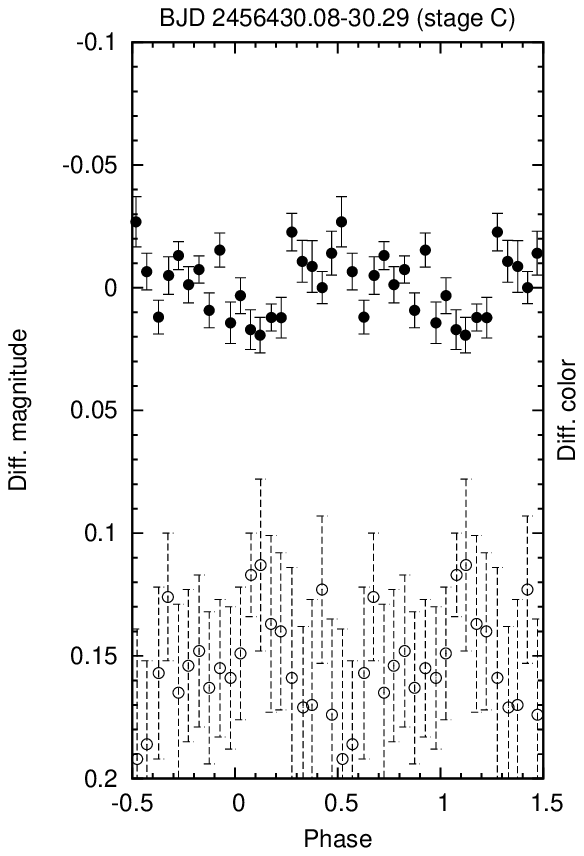}
\FigureFile(40mm,40mm){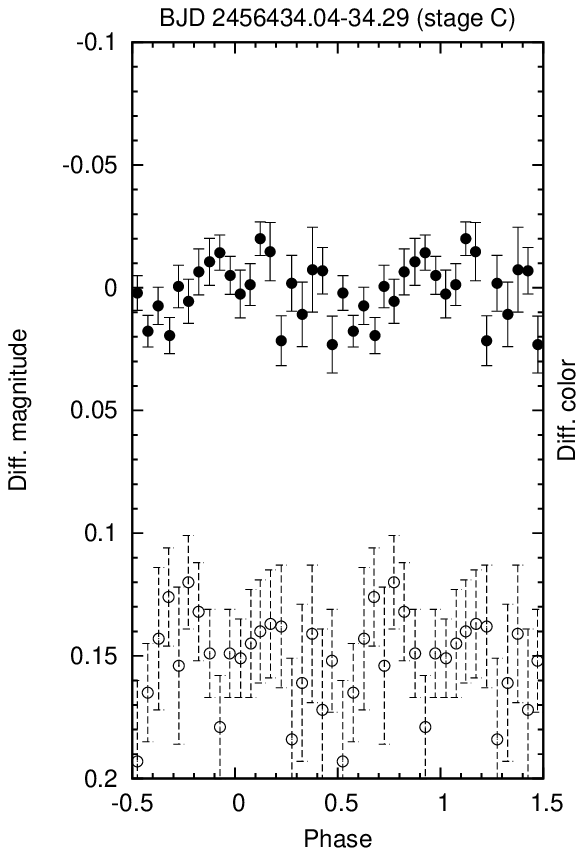}
\FigureFile(40mm,40mm){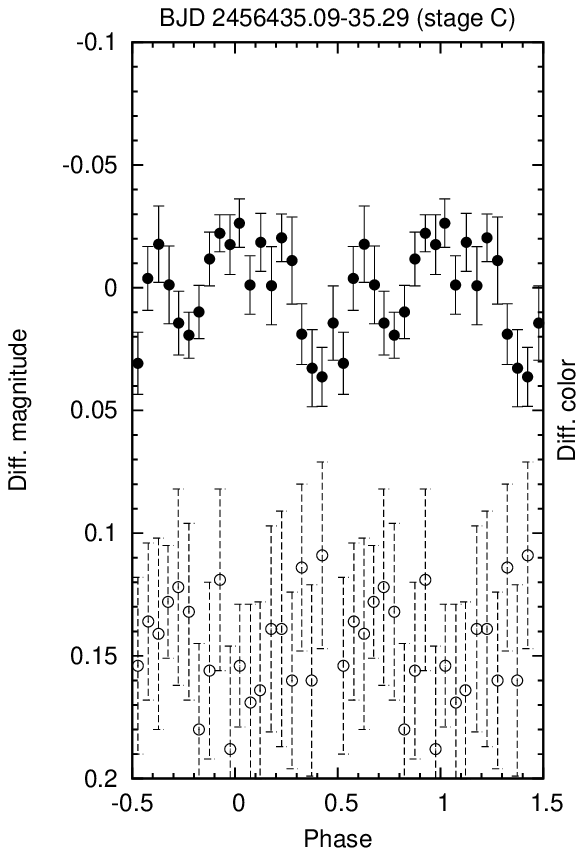}
\FigureFile(40mm,40mm){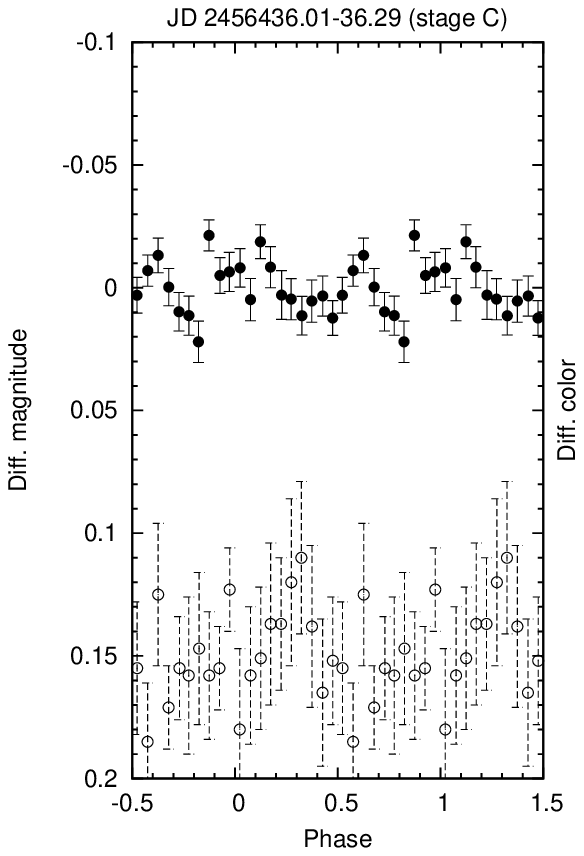}
\end{center}
\caption{Nightly-averaged $R_{\rm c}$ light curves (filled circles) and $g' - I_{\rm c}$ colour variations (open circles) folded with 0.046346 d (stage A), 0.045515 d (stage B), and 0.045473 d (stage C). Single-peaked profiles, characteristic of superhumps, are visible, particularly during the plateau stage. Weak modulations are also visible during the rebrightening stage.
There is a hint that the bluest peaks in $g' - I_{\rm c}$ correspond to the brightness minima in some nights.}
\label{j1740_shvar}
\end{figure*}

\begin{figure*}
\begin{center}
\FigureFile(160mm,160mm){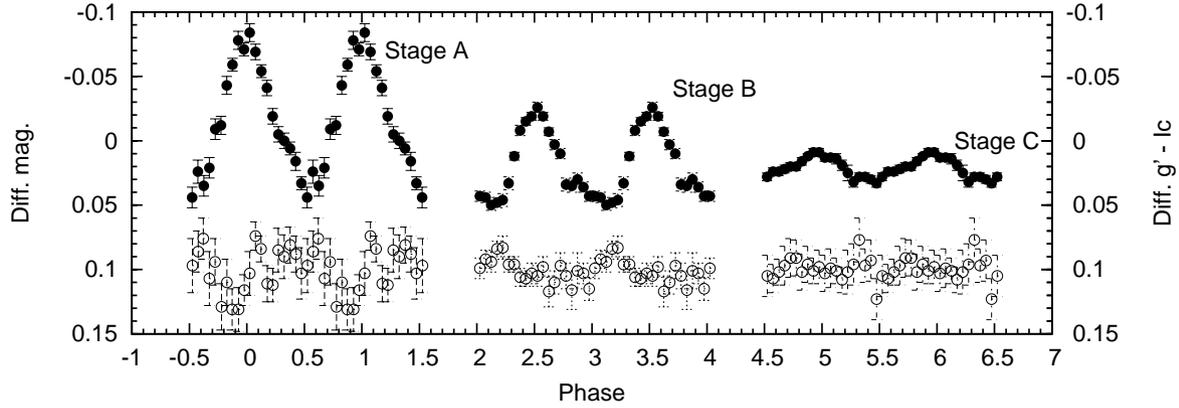}
\end{center}
\caption{$R_{\rm c}$ light curves (filled circles) and $g' - I_{\rm c}$ colour variations (open circles) in each stage folded with 0.046346 d (stage A), 0.045515 d (stage B), and 0.045473 d (stage C). $R_{\rm c}$ light curve is anticorrelated with $g' - I_{\rm c}$ colour variations in stage B. In common, the brightness maxima differ from the bluest peaks in $g' - I_{\rm c}$ colours.}
\label{shvarABC}
\end{figure*}

\begin{figure}
\begin{center}
\FigureFile(80mm,80mm){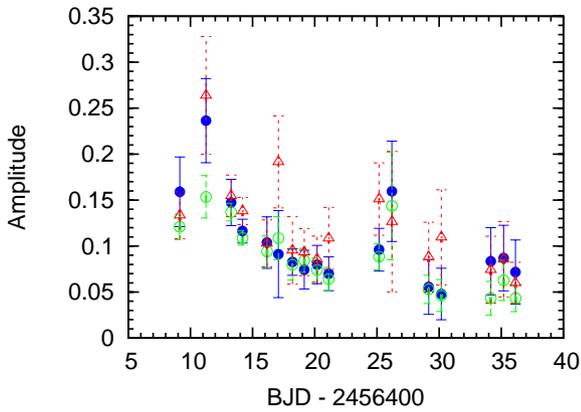}
\end{center}
\caption{Nightly-averaged superhump amplitudes in $g'$ (filled circles, blue), $R_{\rm c}$ (open circles, green), and $I_{\rm c}$ (open triangles, red) bands. The maximum amplitudes occurred on BJD 2456411, corresponding to the beginning of stage B. A regrowth of the superhump amplitudes was observed on BJD 2456425, corresponding to the stage B$-$C transition.}
\label{amp}
\end{figure}

\begin{figure}
\begin{center}
\FigureFile(80mm,80mm){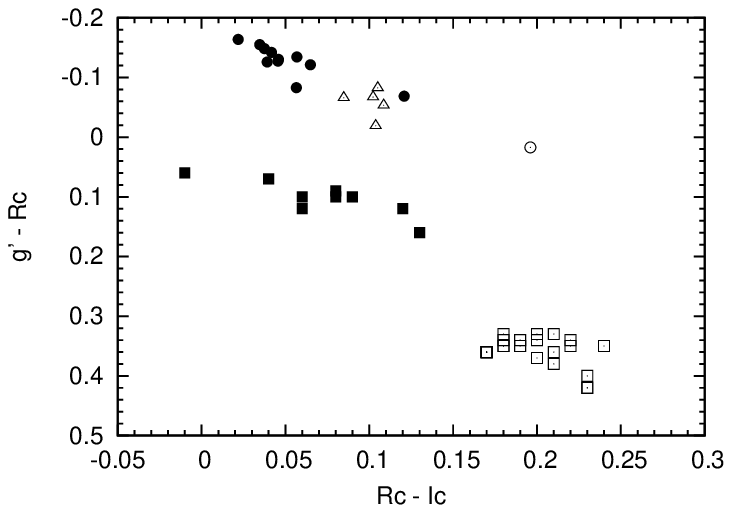}
\end{center}
\caption{Nightly-averaged colour-colour diagram of the superoutburst. The filled circles, open circle, and open triangles indicate the datapoints of the plateau stage, dip, and rebrightening stages of CSS J1740, respectively. The filled and open squares denote the datapoints of the plateau and post-superoutburst stages during the 2007 superoutburst of V455 And. Note that CSS J1740 shows bluer $g' - R_{\rm c}$ colour variations compared with V455 And and the dip of CSS J1740 has a similar $R_{\rm c} - I_{\rm c}$ colour to the post-superoutburst stage of V455 And.}
\label{colcol}
\end{figure}

The superhump maxima are listed in table \ref{o-ctable}. In order to improve our period analyses, we downloaded the AAVSO light curve and combined these with our data. A linear regression to the superhump maxima yields

\begin{equation}
BJD(max) = 2456407.6914(11) + 0.045549(5) \times E.
\end{equation}

Based on this equation, we obtained the $O - C$ diagram of superhump maxima, which is shown in figure \ref{o-c}. As can be seen in the $O - C$ diagram, stage A can be clearly recognized in the early stage of our data. The obtained $O - C$ diagram indicates that stage A ended before BJD 2456409.6 ($E$ = 42). We performed the Phase Dispersion Minimization method (PDM, \citet{pdm}) for stage A superhumps and estimated the mean period as $P$ = 0.046346(67) d. The resultant theta diagram is displayed in the top of figure \ref{pdm}.
The errors in the PDM method are calculated using Lafter-Kinmann class method developed by \citet{fer89error} and \citet{pdot2}.

The $O - C$ diagram also shows that J1740 experienced the stage B-C transition on BJD 2456425. Based on the above analyses, we performed the PDM for stage B (BJD 2456409.60-24.58) and stage C (after BJD 2456425.05), and obtained the best estimated periods to be $P$ = 0.045515(29) d for stage B and $P$ = 0.045473(25) d for stage C, respectively. The resultant theta diagrams are displayed in the middle and bottom of figure \ref{pdm}. These obtained values are in agreement with that derived by \citet{cho15j1740} and \citet{pdot7}. The profile of the $O - C$ diagram means the positive $P_{\rm dot}$ during stage B. By fitting the data between 42$<= E <=$ 370 with a quadratic equation, we obtained $P_{\rm dot}$ to be 1.6 ${\times}$ 10$^{-5}$, in good agreement with that derived by \citet{pdot7}.

\subsection{Superhumps and colour variations}

Figure \ref{j1740_shvar} exhibits nightly-averaged $R_{\rm c}$ light curves and $g'-I_{\rm c}$ colours folded with the above obtained periods. Single-peaked profiles characteristic of superhumps are visible, particularly in the early datasets of our run. Although the profiles became complicated during the rebrightening, hump-like modulations were visible. As for $g'-I_{\rm c}$ colour variations, the bluest peaks tend to coincide with the brightness minima. This tendency is also reported in other WZ Sge-type dwarf novae, such as EZ Lyn \citep{iso15ezlyn}, SSS J122221.7$-$311525 \citep{neu17j1222}, HV Vir, and OT J012059.6$+$325545 \citep{ima18hvvirj0120}.

In order to further study the relation between the superhump and colour variations and compare our data with the previous works, we made phase-averaged $R_{\rm c}$ light curves and $g' - I_{\rm c}$ colour variations in each stage, which are represented in figure \ref{shvarABC}. As noted above, the light minima of superhumps coincide with bluest peaks of $g' - I_{\rm c}$ during stage A and B. There shows a common property that the brightness maximum of superhumps differ from the bluest peak of $g' - I_{\rm c}$ in phase. In recent years, \citet{iso15ezlyn} discussed the values of the phase difference, by using the data of EZ Lyn and V455 And. The phase difference between the superhump light and colour variation was ${\sim}$ 0.5 in EZ Lyn, while that was only ${\sim}$ 0.15 in V455 And (see figure 8 of \citet{mat09v455and}). In the case of CSS J1740, we found ${\sim}$ 0.5 in stage A and 0.3 in stage B, respectively. At present, we cannot draw a firm conclusion to the working mechanism of the phase difference, because of the lack of multicolour photometry during superoutbursts. 
Here we summarize the values of the phase difference in table \ref{offset}. The physical mechanisms of the phase difference should be clarified by collecting further observational samples.

\subsection{Amplitudes of superhumps}

Figure \ref{amp} shows nightly-averaged amplitudes of superhumps in each band. Statistical studies of amplitudes of superhumps have shown that the maximum amplitude of superhumps occurs around the stage A-B transition, and that a regrowth of superhumps occurs around the stage B-C transition \citep{pdot}. As can be seen in this figure, the superhump amplitudes showed the largest values on BJD 2456411, corresponding to the beginning of stage B. A regrowth of the superhumps was also detected on BJD 2456425, corresponding to the beginning of stage C. Judging from these observations, the amplitude development observed in CSS J1740 is common for SU UMa-type dwarf novae.

Recently, \citet{ima18hvvirj0120} reported that the amplitudes of superhumps are likely to be independent of wavelength while those of early superhumps are strongly dependent on wavelength. Although we can find a weak tendency that the $I_{\rm c}$ band has the large amplitudes in stage B and $R_{\rm c}$ band has the small amplitudes in stage C in some nights, these tendencies are not evident compared with that observed in early superhumps in WZ Sge-type dwarf novae \citep{ima18hvvirj0120}.

\subsection{Colour-colour diagram}

In order to study colour development during the superoutburst of CSS J1740, we derived the nightly-averaged colour-colour diagram, which is displayed in figure \ref{colcol}. Also shown are the nightly-averaged colours during the 2007 superoutburst of V455 And, taken from table 2 of \citet{mat09v455and}. At the onset of our observations, the colours of CSS J1740 were located on the top-left region in figure \ref{colcol}, after which the colours became red as the superoutburst proceeded. The colours during the dip significantly deviated from the majority of the datapoints. During the rebrightening stage, the colours were clustered around $g' - R_{\rm c}$ ${\sim}$ -0.05 and $R_{\rm c} - I_{\rm c}$ ${\sim}$ 0.1. We also found that $g' - R_{\rm c}$ colours during the plateau stage of CSS J1740 were about 0.2 mag bluer than those of V455 And, while the $R_{\rm c} - I_{\rm c}$ colours were almost the same between these systems. Because we have fewer samples of the colour-colour diagram during superoutbursts, it is unknown whether CSS J1740 shows bluer $g' - R_{\rm c}$ colour variations than other systems. The textbook colour-colour diagram of superoutbursts should be established. Regarding the optical spectrum, \citet{cho15j1740} reported that CSS J1740 provides evidence for overabundance in helium. This feature is similar to that of the helium-rich dwarf nova SBS J1108+574 \citep{lit13sbs1108}, although helium lines in CSS J1740 are not so strong as in SBS J1108+574. At present, it is unclear whether colour variations of J1740 are affected by a richness in helium, which should be elucidated in future multicolour photometry of various types of dwarf novae.

We also note that the colours of the dip show unique values compared with those of other stages. The dip datapoint of CSS J1740 means that $R_{\rm c} - I_{\rm c}$ colour already returns to the quiescent level, while $g' - R_{\rm c}$ colour remains relatively blue. In order to compare our results with other observations during the dip, we estimated $g' - R_{\rm c}$ and $R_{\rm c} - I_{\rm c}$ colours of the dips during the 1996-97 superoutburst of EG Cnc and 2001 superoutburst of WZ Sge itself. These data were taken from table 2 of \citet{pat98egcnc} and table 2 of \citet{how04wzsge}, and we converted these data to $g' - R_{\rm c}$ by using the equation (1) and (2). Table \ref{dipcol} shows the resultant colours during the dip. The table clearly shows the red colours of the dips for EG Cnc and WZ Sge, while those of CSS J1740 were significantly bluer compared with those of EG Cnc and WZ Sge. Whether or not the blue dip of CSS J1740 is associated with the helium richness should be clarified in further photometric samples during the dip.

\subsection{Mass ratio of CSS J1740}

In recent years, \citet{kat13qfromstageA} have proposed a new method of estimating the mass ratio of the system using the orbital and stage A superhump periods (stage A method). This method is powerful for WZ Sge-type dwarf novae, since the observed early superhump period can be regarded as the orbital period of the system. In addition, \citet{kat15wzsge} has shown that the duration of stage A superhumps tend to be inversely proportional to the mass ratio, which enables us to precisely determine the period of stage A superhumps. Although the late start of our observations prevented us from detecting early superhumps, \citet{cho15j1740} reported that the orbital period of CSS J1740 to be $P_{\rm orb}$ = 0.045028(7) d. By using the analytic formulae described in \citet{kat13qfromstageA} (the exact formulae are described in \citet{nam17}), we estimated the mass ratio of CSS J1740 to be $q$ = 0.077(5). The obtained value is very typical among WZ Sge-type dwarf novae, despite the fact that the orbital period of the system is far below the observational period minimum \citep{kat15wzsge}.

However, our estimation of the mass ratio of CSS J1740 is significantly larger than that derived by \citet{cho15j1740}, who obtained $q$ = 0.0565(20). This discrepancy comes from the fact that
\citet{cho15j1740} used so-called Patterson relation with ${\epsilon}$ = 0.18$q$ + 0.29$q^{2}$, where ${\epsilon}$ is defined as the fractional period excess with $P_{\rm sh}$/$P_{\rm orb}$ - 1 \citep{pat05SH}. Conventionally, the mean $P_{\rm sh}$ was determined by using the total light curve during the plateau stage of the superoutburst, which indicates that $P_{\rm sh}$ is determined by combined stage A, B, and C superhumps. Recent theoretical studies suggest that stage B is the stage that the pressure effect becomes dominant, resulting in changing superhump periods \citep{osa13v344lyrv1504cyg}. Observationally, it is known that $P_{\rm sh}$ of stage A is 1.0$-$1.5\% longer than that of stage B \citep{pdot}. If we determine the mean $P_{\rm sh}$ by using the total light curve during the plateau stage, then the estimated $P_{\rm sh}$ will be very close to $P_{\rm sh}$ during the stage B, since the most of the plateau stage consists of stage B. As a result, the mean $P_{\rm sh}$ may be underestimated, leading to underestimation of $q$ as well, as pointed out by \citet{nak13j2112j2037}. We point out that $q$ derived by \citet{cho15j1740} may be underestimated. The exact mass ratio of CSS J1740, as well as the validity of the stage A method should be confirmed by future spectroscopic observations using a large telescope.

\section{Summary}
In this paper, we report on our analyses of multicolour photometry during the 2013 superoutburst of the short period dwarf nova CSS J1740. We summarize our results as follows:
\begin{itemize}
\item{The superoutburst light curve of CSS J1740 consists of the main plateau stage, temporal dip, and long-lasting rebrightening. The light curve is classified as the type-A rebrightenings of WZ Sge-type dwarf novae.}
\item{In the superhump modulations, there is a tendency that the superhump minima coincide with the bluest peaks of $g' - I_{\rm c}$ colour variations.}
\item{The amplitude development of superhumps is similar to other WZ Sge-type dwarf novae.}

\item{We plot the nightly-averaged colour-colour diagram of CSS J1740 and found that the $g' - R_{\rm c}$ colour indices are bluer compared with those of V455 And.}
\item{The colour indices of the dip was significantly bluer than those of EG Cnc and WZ Sge.}
\item{We obtained $q$ to be 0.077(5). This value is significantly larger than that obtained by \citet{cho15j1740}. We point out that \citet{cho15j1740} may underestimate $q$ because of the usage of the conventional relation between ${\epsilon}$ and $q$.}

\end{itemize}

\begin{ack}
We thank an anonymous referee for helpful comments on the manuscript of the paper. We are grateful to the Catalina Real-time Transient Survey team for making their real-time detection of transient objects available to the public.
We acknowledge with thanks the variable star observations from the AAVSO and VSNET International Database contributed by observers worldwide and used in this research. We thank Dr. Tomohito Ohshima for providing us with information on the object. This work is partly supported by the Publication Committee of the National Astronomical Observatory of Japan (NAOJ). Part of this work is supported by a Research Fellowship of the Japan Society for the Promotion of Science for Young Scientists (KI).
\end{ack}

\bibliographystyle{pasjtest1}

\end{document}